\documentclass[twocolumn,noshowpacs,preprintnumbers,amsmath,amssymb,superscriptaddress]{revtex4}

\usepackage{graphicx}% Include figure files
\usepackage{dcolumn}% Align table columns on decimal point
\usepackage{bm}% bold math

\begin{document}

\title{Single-Dirac-Cone topological surface states on pseudo-IV-VI Semiconductors: Thallium-based III-V-VI$_2$ Ternary Chalcogenides}% Force line breaks with \\

\date{9 March, 2010}% It is always \today, today,

\author{H. Lin}
\affiliation{Department of Physics, Northeastern University, Boston, Massachusetts 02115, USA}
\affiliation{Joseph Henry Laboratories of Physics, Princeton University, Princeton, New Jersey 08544, USA}
\author{R.S. Markiewicz}
\affiliation{Department of Physics, Northeastern University, Boston, Massachusetts 02115, USA}
\author{L.A. Wray}
\affiliation{Joseph Henry Laboratories of Physics, Princeton University, Princeton, New Jersey 08544, USA}
\affiliation{Princeton Center for Complex Materials, Princeton University, Princeton, New Jersey 08544, USA}
\author{L. Fu}
\affiliation{Department of Physics, Harvard University, Cambridge, Massachusetts 02138, USA}
\author{M.Z. Hasan}
\affiliation{Joseph Henry Laboratories of Physics, Princeton University, Princeton, New Jersey 08544, USA}
\affiliation{Princeton Center for Complex Materials, Princeton University, Princeton, New Jersey 08544, USA}
\author{A. Bansil}
\affiliation{Department of Physics, Northeastern University, Boston, Massachusetts 02115, USA}
\email{bansil@neu.edu}

\begin{abstract}
We have investigated several classes of strong spin-orbit chalcogenides related to the (Pb,Sn)Te series studied in connection with the Dirac fermion physics in the 1980s. Our first-principle theoretical calculations suggest that ternary chalcogenides TlBiX$_2$ and TlSbX$_2$ (X=Te, Se, S) series harbor small bandgap topological insulators with single Dirac cone on some selective surfaces whereas the isostructural and isoelectronic silver-based AgBiX$_2$ and AgSbX$_2$ (X=Te, Se, S) series do not. We find that several Tl-compounds are in the vicinity of a topological critical point. We identify the precise surface termination that realizes the single Dirac cone. The single-Dirac-cone surface is shown to be correlated with a termination that minimizes dangling bonding effects favorable for ARPES experiments. Our results further suggest this class of topological semi-metals may harbor odd-parity topological superconductors similar to the possibility proposed for Cu$_x$Bi$_2$Se$_3$ (T$_c$ $\sim$ 4K).

\end{abstract}

\maketitle

Topological insulators are a recently discovered new phase of quantum matter \cite{moore, hasankane, zhang}.
The search for topological insulators in real materials has benefitted from the fruitful interplay between topological
band theory and realistic band structure calculations \cite{HgTe, fukane, bisezhang, bisb, bisb2, bise, bite, bitedavid, davidprl}.  As a result, a number of materials Bi$_x$Sb$_{1-x}$, Bi$_2$Se$_3$ and
Bi$_2$Te$_3$ have been experimentally realized as three-dimensional topological insulators\cite{bisb, bise, bisb2, bite, bitedavid, davidprl}. Recently the search for topological insulators has been extended to ternary compounds \cite{heuslerhasan, heuslerzhang}. In this work, we report first-principles calculations of (Tl/Ag)-based III-V-VI$_2$ ternary chalcogenide series, and compare their band structures with those of the related (Pb,Sn)Te series studied previously in connection to Dirac fermion physics \cite{fradkin}. We have also carried out slab calculations to study the nature of the surface states. We find a specific surface termination which gives rise to simple Dirac-cone topological surface states due to the non-trivial band topology. We also  propose several ways to engineer the band structure of thallium-based chalcogenides towards achieving the single-Dirac-cone topological insulator phases.

\begin{figure}
\includegraphics[width=0.5\textwidth]{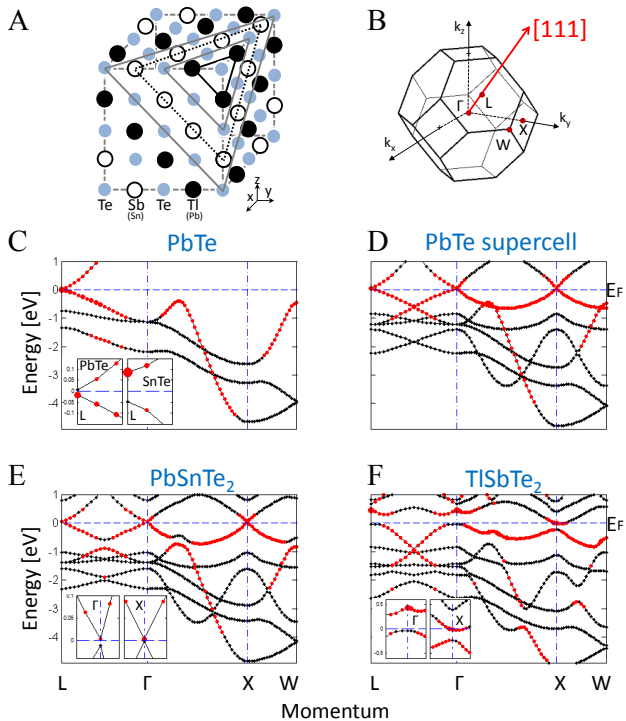}% Here is how to import EPS art
\caption{\label{fig:sketch}
\textbf{Topological band-structure of Tl(Sb/Bi)Te$_2$} : A. Crystal structure of an idealized TlSbTe$_2$ with face-centered-cubic lattice.  Tl, Sb, and Te are denoted by black, open, and gray circles, respectively. The Tl surface is emphasized by black lines, the Sb surface by dotted lines, and the Te surface by gray lines.
B. The Brillouin zone of a FCC lattice. The red line indicates the 111 direction for rhombohedral distortion.
C-F the band structure of PbTe, PbTe calculated in rhombohedral structure with 4 atoms in one unit cell, PbSnTe$_2$, and TlSbTe$_2$. The FCC symbols for points in the Brillouin zone are used and defined in panel B for ease of comparison.
The size of the red dots denotes the probability of s-orbital occupation on the $M$ atoms. The inset highlights the band inversion which did not occur in panels C and D but does occur in panels E and F.
}
\end{figure}

Designing new topological insulators involves modifying their structure or doping to shift band orders out of the natural atomic sequence.
As an example, we consider the well-known case of (Pb,Sn)Te with rocksalt structure.  The end phase PbTe,which possesses a face-centered cubic (FCC) lattice, is topologically trivial. Relative to PbTe, SnTe has band inversions at four equivalent $L$-points in such a way that the parity values of conduction and valence band are switched (see insets in Fig.~1C). Since the inversion occurs at an even number of points in the Brillouin zone, SnTe is also a topologically trivial band insulator.
Fu and Kane\cite{fukane} proposed that a rhombohedral distortion along a particular 111 direction can bring (Pb,Sn)Te to a strong topological phase. This works since the band inversion occurs only at $L$-point along the 111 direction which is distinguished from the other three $L$-points (Fig.~\ref{fig:sketch}B).

The thallium-based III-V-VI$_2$ ternary chalcogenides $MM'X_2$ are or can be approxminately viewed as a rhombohedral structure with the space group $R\bar{3}m$\cite{ThalliumPRB} which possesses real-space inversion symmetry. The unit cell contains four atoms. The $M$ = Tl, $M'$ = Bi or Sb, and $X$ = Te, Se, or S atoms occupy the Wyckoff 3a, 3b, and 6c positions respectively. If $M$ = $M'$ and the corresponding hexagonal lattice constants a and c have the relation c = 2$\sqrt{6}$a, cubic symmetry is restored and the structure can be described as a rocksalt FCC lattice with a$'$ = $\sqrt{2}$ a which contains only two atoms per unit cell. The rhombohedral lattice is embedded in a 2x2 supercell of the FCC lattice as shown in Fig.~\ref{fig:sketch}A. This relation between rhombohedral and FCC lattices is analogous to the type II antiferromagnet NiO, in which case $M$ = Ni spin up, $M'$ = Ni spin down, and $X$ = O. The thallium-based III-V-VI$_2$ ternary chalcogenides can also be considered as an offshoot of IV-VI semiconductors, and therefore referred to as pseudo IV-VI semiconductors\cite{pseudoPbTe, tlbite2structure, Tlmodel, TlBiTe2inversion}. Let us take TlBiTe$_2$ as an example. Because the elements Tl and Bi precede and follow Pb in the periodic table, TlBiTe$_2$ resembles PbTe where $M$ = $M'$ = Pb. In brief, by distorting the PbTe crystal along a 111 direction and replacing Pb atoms by Bi and Tl atoms alternating along that 111 direction, one obtains TlBiTe$_2$ with rhombohedral structure.

Since thallium-based III-V-VI$_2$ ternary chalcogenides provide a rhombohedral version of (Pb,Sn)Te, one may expect that they also can display topologically non-trivial phases.  To explore this issue, first-principles band calculations were performed with the linear augmented-plane-wave (LAPW) method using the WIEN2K package\cite{wien2k} in the framework of density functional theory (DFT). The generalized gradient approximation (GGA) of Perdew, Burke, and Ernzerhof\cite{PBE96} was
used to describe the exchange-correlation potential.  Spin orbital coupling (SOC) was included as a second variational step using scalar-relativistic eigenfunctions as a basis. The radius of the atomic spheres is 2.5 Bohr. The lattice constant were taken from the optimized lattice constants from table I of Ref.~\onlinecite{ThalliumPRB}.

To better illustrate the relation between (Pb,Sn)Te and thallium-based III-V-VI$_2$ ternary chalcogenides, In Fig.~1 we show band dispersions step-by-step for a sequence of structures which systematically evolve from FCC PbTe (Fig.~\ref{fig:sketch}C) to TlSbTe$_2$ (Fig.~\ref{fig:sketch}F). For PbTe in the FCC lattice, the top of the valence bands is located at the L-points and a small gap is found. We then repeat the same calculation, but assuming a rhombohedral unit cell which contains 4 atoms or two chemical formula units, doubled along a particular 111 axis, to obtain the band structure in Fig.~\ref{fig:sketch}D. As the unit cell is doubled in size, the bands are folded. In particular, the top of the valence bands at one L-point is folded to the $\Gamma$-point, and those at the other three $L$-points are folded to three equivalent $X$ points. Note that at this stage the supercell is introduced as a purely formal device: since no distortions are introduced into the structure yet, the bands have exactly the same dispersion and do not interact with the shadow bands introduced by folding.  In the next step (Fig.~\ref{fig:sketch}E), we replace one of the Pb atoms by Sn atom,  resulting in alternating Pb and Sn planes along the selected 111 direction. The dispersion of the original and folded bands now interact, leading to new band gaps and avoided crossings for the hypothetical ternary compound PbSnTe$_2$. Otherwise, the band structure superficially looks quite similar to that found in Fig.~\ref{fig:sketch}D. However, we now show that the conduction and valence bands have become inverted at high symmetry points after the Pb-to-Sn replacement.
To monitor band inversion, we pay attention to the probability of s-orbital occupation on the $M$ atom (here $M$ = Pb) at inversion center, whose magnitude is denoted by red dots. Since the s-orbital occupation must vanish for odd-parity band
at high symmetry points, we use it as an indicator for band parity.
Comparing the band sequence with red dots for PbTe and PbSnTe$_2$, we find that band inversion indeed occurs at both the $\Gamma$-point and the $X$-points, most likely due to a weaker SOC and a smaller atomic size in Sn as compared to Pb.
Because of an even number of band inversions, PbSnTe$_2$ is also topologically trivial similar to SnTe.
However, since PbSnTe$_2$ only has rhombohedral instead of cubic symmetry, its band gaps at the $\Gamma$ and $X$-points become unequal.
The band gaps at $X$ points were found to be smaller, which indicates band inversion from PbTe to PbSnTe$_2$ occurs first at $\Gamma$ and then at $X$.
This raises the hope that by introducing a small rhombohedral lattice distortion or replacing Pb and Sn with other chemical elements, the second band inversion associated with the three $X$-points might be removed, leaving only a single band inversion relative to PbTe (in supercell) at the $\Gamma$-point which leads to a topologically non-trivial band structure.
Based on the close relations between (Pb, Sn)Te and TlM'Te$_2$, we therefore speculate that ternary compounds TlM'Te$_2$ may have the desired band inversion described above.
To test this idea, we use the optimized lattice constants from table I of Ref.~\onlinecite{ThalliumPRB}, to generate the band structure of
TlSbTe$_2$ shown in Fig~\ref{fig:sketch}F. The overall band structure resembles that of (Pb,Sn)Te.
The rhombohedral lattice distortion enhances all band gaps, especially at $\Gamma$, and shifts the band edges at the $X$-points. Our calculations shows that the  bottom of the conduction band at the $X$-point overlaps the top of the valence band at a low symmetry point (not shown), leading to a semi-metal ground state with small electron pockets at the $X$-points. Nevertheless, since a direct band gap between the conduction and valence bands exists at every point in the Brillouin zone, the $Z_2$ topology of valence bands are well defined.
To study the band topology, one should focus on the band inversion relative to PbTe (in supercell) at the $\Gamma$-point and three $X$-points.
As can be seen from the s-orbital occupation in Fig.~1F, band inversion occurs at all four points, as in PbSnTe$_2$. So according to our DFT-GGA calculations,
 TlSbTe$_2$ is a topologically trivial semi-metal.

\begin{figure}
\includegraphics[width=0.5\textwidth]{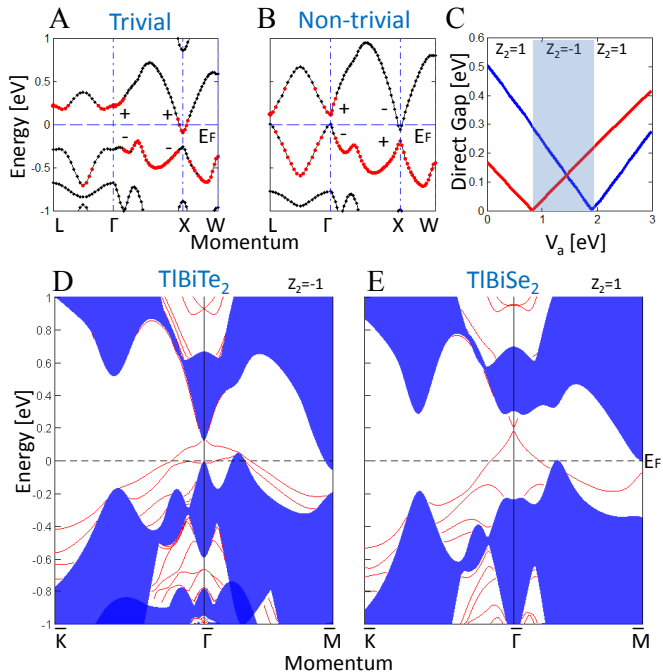}% Here is how to import EPS art
\caption{\label{fig:bulkbands} \textbf{Band inversion and Parity analysis}: Panels A and B show the bulk band structure of TlBiTe$_2$ without and with bandgap corrections, respectively. FCC notation for the high-symmetry points of the Brillouin zone is used and defined in Fig.~1B for ease of comparison. The band gap as a function of added orbital dependent potential $V_a$ is shown in C. The gray area highlights the features related to topological non-triviality. D and E show surface state dispersions in red lines. The projected bulk bands are shown as blue areas.
}
\end{figure}

Figure~2(A) illustrates the band structure of another member of this family, TlBiTe$_2$.  It is seen to also be a topologically trivial semi-metal, as are all six thallium-based III-V-VI$_2$ ternary chalcogenides presented here, according to DFT-GGA calculations.  However, it is a well known problem of DFT that the band gap in semiconductors is underestimated. The simplest way to correct the band gap is the scissor operation that shift rigidly the conduction band toward higher energy. Here, because band inversion necessarily involves orbital hybridization, such a rigid shift of the conduction band cannot be directly applied. As an alternative, we shift the conduction band by adding orbital dependent potentials into the Hamiltonian, allowed in a full-potentials code such as Wien2k (this procedure underlies the well-known LDA+U method). This allows us to test the robustness of the band topology against band gap corrections. Applying the orbital potentials on one specific atom can also be considered as changing the local potential on site. This allows us to test the robustness of band topology against alloying.
For thallium-based III-V-VI$_2$ ternary chalcogenides, the orbital character of the valence band is p-type of $X$ atoms, while the conduction band is primarily p-type of $M$' atoms. We then add orbital dependent potentials $V_a$ to all three p-orbitals of $M$' to mimic the band gap correction or alloying on this site which moves the conduction band upwards. For an example, we study TlBiTe$_2$, Fig.~2(A). DFT-GGA predicts TlBiTe$_2$ to be a topologically trivial metal due to an even number of band inversion.
As $V_a$ increases, the whole conduction band move upwards, but the direct gaps at both $\Gamma$ and $X$-points become smaller.
As shown in Fig.~2C, the gap at $X$ first vanishes at $V_a$=0.8eV at which the conduction band and valence band join and form a 3D Dirac-cone.
The transition for the $\Gamma$ point occur at larger value 1.9eV of $V_a$.
In between these two critical points, bands are inverted only at the $\Gamma$-point relative to PbTe and the system becomes topologically non-trivial (gray area in Fig.2C), with $Z_2$ topological index $(1;000)$. Similar band inversions at $X$-points after band correction have been found in other compounds of this family as well. Despite uncertainties inherent in first-principles electronic structure calculations, we have identified the band inversion at $X$-points (relative to PbTe) as the decisive factor of their topological class, which by itself suggests several ways to engineer the band structure towards the topological insulator side, including rhombohedral distortion and alloying.

A non-trivial band topology will generate metallic surface states which are the hallmark of topological insulators. For the 111 surface in thallium-based III-V-VI$_2$ ternary chalcogenides, the atoms are sequenced as -$M$-$X$-$M'$-$X$- and there are four possible surface terminations (see Fig.~1A).
Unlike in topological insulators Bi$_2$Se$_3$ and Bi$_2$Te$_3$, the bonding between the layers in TlM'X$_2$ is not weak and there are no natural cleavage planes. As a result, the topologically protected surface states may coexist with non-topological ones arising from dangling bonds, leading to a complicated surface spectrum. We have carried out extensive slab calculations on all possible surface terminations for all six compounds to search for a simple surface spectrum. We have found that the termination between $X$ and $M$ atoms can give a such case. On the $X$ exposed surface, there are much less trivial surface states compared with other surface terminations.
%We also speculate that it is the most likely surface termination since the breaking of the bonding on the surface make %less influence on the band structure.
In Fig.~2(D), we show the surface band structure of TlBiTe$_2$, assuming $V_a$ = 1.5 eV to be in the topological non-trivial phase. The surface states are shown as red lines. The calculation is based on a slab with 47 atomic layers. Between $\Gamma$ and $M$ points, there is only one surface band connecting conduction band and valence band. This is the unambiguous evidence of non-trivial topological nature.

\begin{figure}
\includegraphics[width=0.5\textwidth]{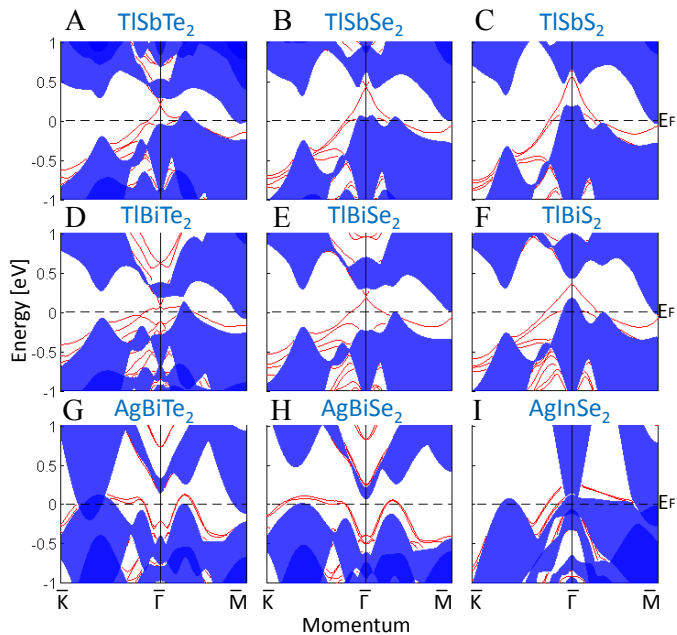}% Here is how to import EPS art
\caption{\label{fig:HgTe}
Band structures based on thin slab calculations for TlSbTe$_2$ (A), TlSbSe$_2$ (B), TlSbS$_2$ (C), TlBiTe$_2$ (D), TlBiSe$_2$ (E), TlBiS$_2$ (F), AgBiTe$_2$ (G), AgBiSe$_2$ (H), and AgInSe$_2$ (I) without bandgap corrections. The blue area indicates the projected bulk bands. The red lines are surface states.
}
\end{figure}

Even though all six compounds are found to be topologically trivial by DFT-GGA calculations without bandgap correction, their surface states may have Dirac-cone dispersions due to the band inversion. In Fig.3, we show the results without bandgap corrections for 23 atomic layer slab where the $X$ atoms are exposed at both sides of the slab.
The thickness is about 50 Angstroms. A Dirac cone with a small gap at $\bar{\Gamma}$ is obtained in TlSbTe$_2$ (A), TlSbSe$_2$ (B), TlBiTe$_2$ (D), and TlBiSe$_2$ (E). The gap is found to decrease rapidly as the slab thickness increases, indicating that a gapless Dirac band exists at $\Gamma$ on the surface of the infinite thickness limit.
Therefore the existence of a surface Dirac band at $\Gamma$ by itself cannot reveal the band topology.
This indicates that a k $\cdot$ p model Hamiltonian around $\Gamma$ is not adequate for these compounds.
Since band topology is determined by the band structure near $X$-points,
it is the surface spectrum near $\bar{M}$-points that distinguishes trivial and non-trivial phases.
In our slab calculations, surface states near $\bar{M}$ have strong finite size effects arising from interactions between the two sides of the slab. This is a consequence of the smaller band gap near $\bar{M}$ found in our bulk calculations. We note that AgBiTe$_2$, AgBiSe$_2$, and AgInSe$_2$ all have a similar crystal structure to the present compounds, but with $M$ = Ag. However, there is no band inversion from our calculation. The surface states of the thin-film are just pull out from the bulk bands. They have similar band dispersion as bulk bands. There is no protected Dirac cone like surface band.

Apart from the fact that Tl-compounds are usually semi-metallic or very weakly semiconducting, they are topologically very similar to the Bi$_2$Se$_3$ series discovered previously. Recently, it was discovered that upon Cu doping, Bi$_2$Se$_3$ becomes a superconductor with a T$_c$ = 3.8K and exhibit unconventional band topology \cite{hor, wray, toposc}. The doped topological states in the Tl-based compounds are also likely to become superconductors. The Tl-series thus provides a fertile ground to search for odd-parity topological superconductors candidates similar to the Cu$_x$Bi$_2$Se$_3$ series \cite{hor, wray, toposc}.

In conclusion, we have shown that thallium-based III-V-VI$_2$ ternary chalcogenides harbor a band-structure matrix for Z$_2$ =-1 topological states and unlike the Bi$_2$Se$_3$ series the topological state in the Tl-chalcogenides critically depend on the nature of the band ordering near the $X$ point of the Brillouin zone and the topological nature is shown to be characterized by band inversions at both $\Gamma$ and $X$-points. We identified the material surface termination that hosts the single Dirac cone surface states. The overall results suggest that these materials can be a new platform to search for topological superconductors along the lines realized in Cu$_x$Bi$_2$Se$_3$ \cite{hor, wray, toposc}. 
After the completion of this work, we became aware of a preprint by Yan et al. \cite{thalliumzhang} on the present Tl-compounds. In contrast, however, we have identified through first principles computations the specific surface termination plane that features the protected single Dirac cone suitable for photoemission (ARPES) studies of this class of materials. Moreover, we have clarified the relationship of the Tl-compounds with PbTe, obtaining insight into pathways for engineering topologically interesting properties.

We acknowledge discussions with R.J. Cava, B.A. Bernevig and C.L. Kane. This work is supported by Basic Energy Sciences of U.S. DOE, Northeastern, Princeton and Harvard University.


\begin{thebibliography}{99}

\bibitem{moore}
J. E. Moore, Nature \textbf{464}, 194 (2010).

\bibitem{hasankane}
M. Z. Hasan and C. L. Kane, arXiv:1002.3895 (2010).

\bibitem{zhang}
X.-L. Qi and S.-C. Zhang, Physics Today \textbf{63}, 33 (2010).

\bibitem{HgTe} B.A. Bernevig {\it et al.}, Science \textbf{314}, 1757 (2006).

\bibitem{fukane}
L. Fu and C.L. Kane, Phys. Rev. B {\bf 76}, 045302 (2007).

\bibitem{bisb}
D. Hsieh {\it et al.},
% D. Qian, L. Wray, Y. Xia, Y. S. Hor, R. J. Cava, and M. Z. Hasan,
Nature (London) {\bf 452}, 970 (2008).

\bibitem{bisb2}
D. Hsieh {\it et al.},
% D. Qian, L. Wray, Y. Xia, Y. S. Hor, R. J. Cava, and M. Z. Hasan,
Science {\bf 323}, 919 (2009).

\bibitem{bise}
Y. Xia {\it et al.}, Nature Phys. {\bf 5}, 398 (2009).

\bibitem{bisezhang}
H. Zhang {\it et al.}, Nature Phys. {\bf 5}, 438 (2009).

\bibitem{bite}
Y.L. Chen {\it et al.}, Science {\bf 325}, 178 (2009).

\bibitem{bitedavid}
D. Hsieh {\it et al.}, Nature (London) {\bf 460}, 1101 (2009).

\bibitem{davidprl}
D. Hsieh {\it et al.}, Phys. Rev. Lett. \textbf{103}, 146401 (2009).

\bibitem{heuslerhasan}
H. Lin {\it et al.},  arXiv:1003.0155 (2010).

\bibitem{heuslerzhang}
S. Chadov {\it et al.}, arXiv:1003.0193 (2010).

\bibitem{fradkin}
E. Fradkin {\it et al.}, Phys. Rev. Lett. \textbf{57}, 2967 (1986).

\bibitem{wien2k}
%
P. Blaha, K. Schwarz, G. K. H. Madsen, D. Kvasnicka, and J. Luitz, WIEN2k, An Augmented Plane Wave Plus Local Orbitals
  Program for Calculating Crystal Properties, (Karlheinz Schwarz, Techn. University Wien, Austria), 2001.ISBN 3-9501031-1-2.

\bibitem{PBE96}
P. Perdew \emph{et al.,} Phys. Rev. Lett.\textbf{ 77}, 3865 (1996).

\bibitem{Wyckoff}
R.W.G. Wyckoff, Crystal Structures (Krieger, Melbourne, FL, 1986), Vol. 2.

\bibitem{FuPRB}
L. Fu and C. L. Kane, Phys. Rev. B \textbf{76}, 045302 (2007).

\bibitem{ThalliumPRB}
K. Hoang and S. D. Mahanti, Phys. Rev. B \textbf{77}, 205107 (2008).

\bibitem{pseudoPbTe}
K. M. Paraskevopoulos, J. Phys. C: Solid State Phys. \textbf{18}, 4941-4949 (1985).

\bibitem{tlbite2structure}
O. Valassiades {\it et al.}, Phys. Stat. Sol. (a) \textbf{66}, 215 (1981).

\bibitem{Tlmodel}
D. V. Gitsu {\it et al.}, J. Phys.: Condens. Matter \textbf{2}, 1129-1140 (1990).

\bibitem{TlBiTe2inversion}
K. M. Paraskevopoulos {\it et al.}, Journal of Alloys and Compounds \textbf{467}, 65 (2009).

\bibitem{hor} Y.S. Hor {\it et al.}, Phys. Rev. Lett. 104, 057001 (2010).
\bibitem{wray} L.A. Wray {\it et al.}, arXiv:0912.3341v1 (2009).

\bibitem{toposc} L. Fu and E. Berg arXiv:0912.3294v1 (2009).

\bibitem{thalliumzhang}
B. Yan {\it et al.}, arXiv:1003.0074 (2010).


\end{thebibliography}
\end{document}